\documentclass[aps,pre,twocolumn]{revtex4}

\usepackage{graphicx}
\usepackage{amsmath}

\begin{document}
\title{Cavity Optomechanics of Levitated Nano-Dumbbells:\\
Non-Equilibrium Phases and Self-Assembly }

\author{W. Lechner$^{1,2}$}
\email{w.lechner@uibk.ac.at}
\author{S.~J.~M.~Habraken$^{1,2}$}
\author{N. Kiesel$^{3}$}
\author{M. Aspelmeyer$^{3}$}
\author{P. Zoller$^{1,2}$}

\affiliation{$^1$Institute for Quantum Optics and
Quantum Information, Austrian Academy of Sciences, 6020 Innsbruck, Austria}
\affiliation{$^2$Institute for Theoretical Physics,  University of Innsbruck,
6020 Innsbruck, Austria}
\affiliation{$^3$Faculty of Physics, University of Vienna, Boltzmanngasse 5, 1090 Vienna, Austria}

\date{\today}

\begin{abstract}
Levitated nanospheres in optical cavities open a novel route to study many-body systems out of solution and highly isolated from the environment. We show that properly tuned optical parameters allow for the study of the non-equilibrium dynamics of composite nano-particles with non-isotropic optical friction. We find optically induced ordering  and nematic transitions with non-equilibrium analogs to liquid crystal phases for ensembles of dimers. 
\end{abstract}

\maketitle
\textit{Introduction -} The interaction between light and matter has been one of the central driving forces behind recent developments in condensed matter physics with nanoparticles \cite{SOFTMATTERREVIEW}. Optical tweezers \cite{ASHKIN,GRIER} and confocal microscopy \cite{CONFOCAL} have made it possible to study many-body systems of nanoparticles in solutions in real-time and with single-particle resolution \cite{SCIENCE_GASSER,MARET,BECHINGER,PERTSINIDIS_PRL}. Recently, it has been proposed to optically levitate and cool single nanospheres inside an optical cavity \cite{CHANGZOLLER,CIRAC,LARGEQUANTUM,CKLAW}. While in the realm of soft-matter physics, this may provide an alternative to confinement in a solution \cite{BROWNIAN}, from a quantum-optics point of view such a set-up provides a versatile alternative to conventional optomechanical systems \cite{KIPPENBERG,HARRIS,REVIEWOPTOMECHANICS}. Combined with optomechanical cooling and trapping techniques of single particles, this may even open the possibility to study fundamental aspects of quantum mechanics with mesoscopic objects \cite{PLANCKSCALE,LARGEQUANTUM,SHORTWALK}. Here, we focus on the dynamics of many, interacting particles in the presence of optomechanical cooling. While many-body systems with non-uniform cooling have been studied with atoms and ions \cite{IONREVIEW}, the possibility to create complex structures with nanospheres offers completely new opportunities to study pattern formation and self assembly. With novel synthesis methods it is now possible to  design compound structures ranging from dimers to networks of nanospheres connected by spring-like biomolecules \cite{DNADUMBBELL, DNABUSTAMANTE,DNATEMPLATING,PINE,DNATEMPLATING2,COLLOIDSDUMBBELLS, ALIVISATOS}. A distinctive feature of the self-assembly of composite particles is that the emerging patterns are characterised not only by their positions, but also by their individual orientations \cite{SMECTIC,SOFTMATTERBOOK}. The non-equilibrium self-assembly of such nano-structures in the presence of non-isotropic optical cooling is an open question, and holds the promise of new means to optical control of pattern formation and novel non-equilibrium liquid crystal phases.

\begin{figure}[hb]
\centerline{\includegraphics[width=8.5cm]{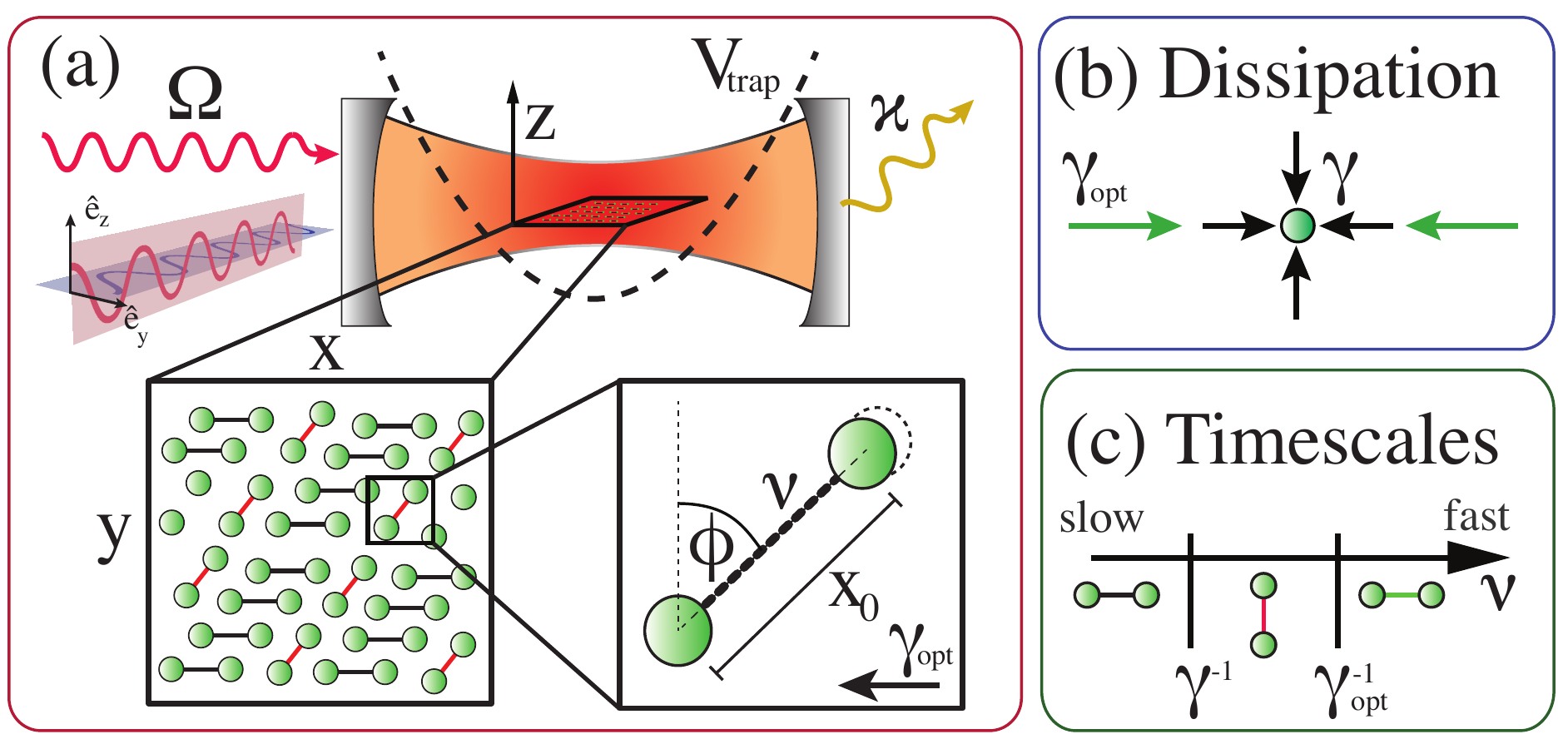}}
\caption{(a) Levitated dimers composed of dielectric nano-spheres in a laser driven optical two-mode cavity with laser strengths $\Omega_{1,2}$, a decay rate $\kappa$ and an additional harmonic trap $V_{\rm trap}$. The orientation of the dimers is denoted $\phi$, their equilibrium separation $x_0$ and their frequency $\nu$. The direct pair interaction between the nanospheres leads to liquid crystal phases with rich dimer patterns. (b) The particles are in contact with a thermal bath with coupling $\gamma$. The cavity-particle interaction results in strong additional dissipation in one spatial direction only, which acts as an effective zero temperature bath $\gamma_{\mbox{opt}}$. (c) Comparison of the relevant time scales: We assume that $\gamma_{\rm opt}$ is faster than $\gamma$ and find $3$ qualitatively different regimes, which give rise to different orientations, depending on the relative timescales. 
}
\label{fig:illustration}
\end{figure}
\begin{figure}[hb]
\centerline{\includegraphics[width=7cm]{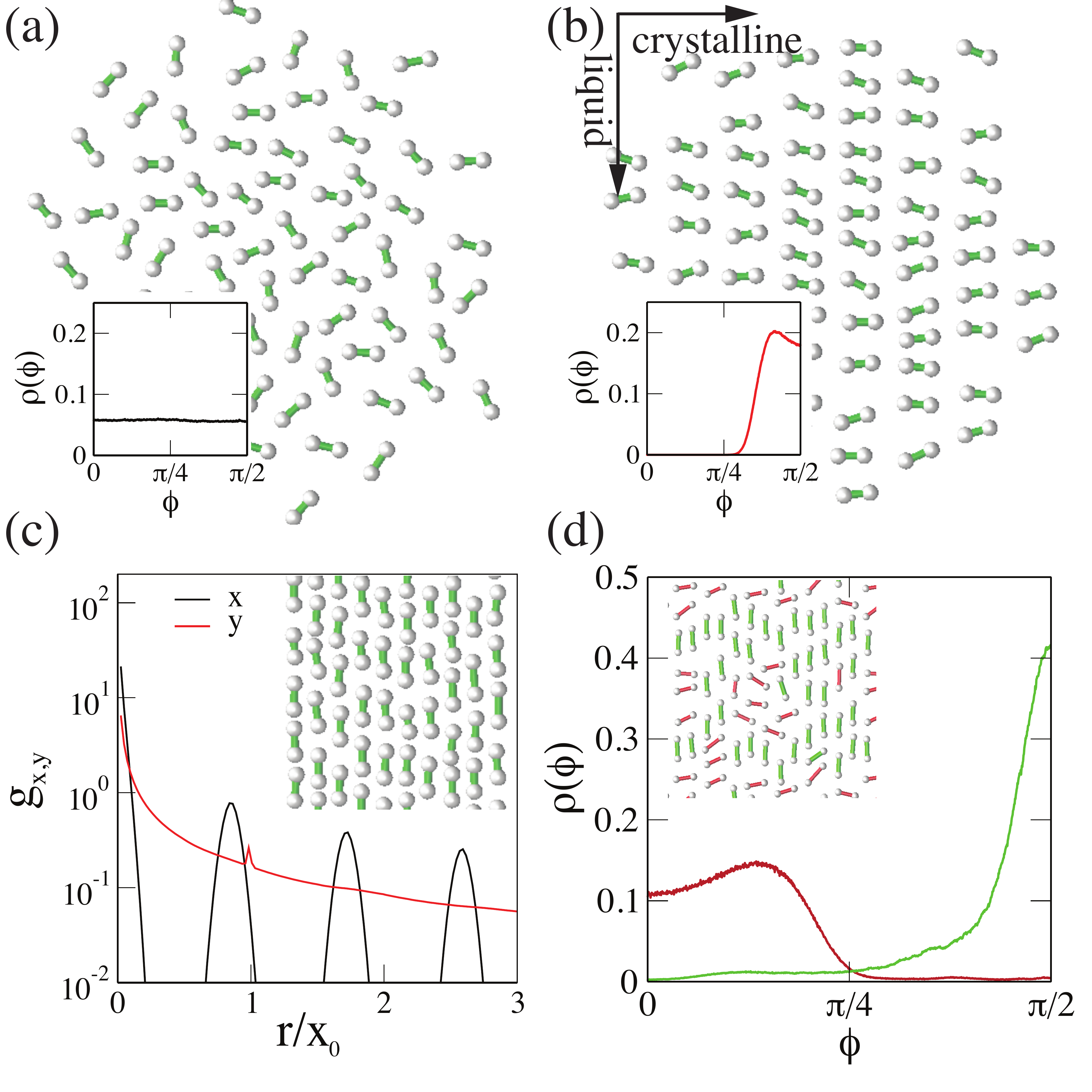}}
\caption{(a) Ensemble of dimers with frequency $\nu=200 \gamma$ (intermediate regime (ii)) in thermal equilibrium are uniformly orientated (inset a). (b) The non-isotropic optical friction induces orientational ordering of individual dimers (inset b) and in addition, many-body nematic ordering along the direction of cooling. (c) The non-equilibrium phase analog to a liquid crystal phase characterized by the pair-correlation function $g(x)$. The rigid rotors align orthogonal to cooling with additional nematic order in direction of cooling (black) and liquid order in $y$-direction (red). (d) Mixture of dimers self assembles into a crystal with dimer-orientations according to the frequencies $\nu$. }
\label{fig:liquidcrystal}
\end{figure}

In this Letter, we study the dynamics and self-assembly of levitated nanosphere-dimers in presence of optomechanical friction inside a two-mirror cavity. The particles are subject to thermal forces and coupled to a cavity mode which is driven by an external laser and damped by the cavity-decay. The optomechanical interaction gives rise to an optical potential and a cooling force along the cavity axis. By compensating the potential with a second optical mode, the remaining optomechanical effect is friction along the cavity axis. Fig.~\ref{fig:illustration} illustrates the system we have in mind. The nanospheres are harmonically trapped and confined to the xy plane, where $x$ is the cavity axis. We show that the steady-state orientation of a single dimer is non-uniform in presence of optical friction. Remarkably, the full range of preferred orientations from $0$ to $\pi/2$ is accessible by appropriately tuning the experimental parameters. In a many-body system, the presence of additional direct interactions between the individual nanospheres, leads to competition between the natural triangular ordering of a two-dimensional crystal and dissipation-induced ordering. Compared to other approaches to dynamical ordering of dimers, such as shear \cite{SHEAR} or static electric fields \cite{ELECTRICORDERING}, this method offers additional advantages: (a) the orientation depends on the frequency of the vibrational mode of the dimers which allows for individual ordering in multi-species systems and (b) in addition to ordering at the level of single particles non-uniform friction also leads to novel liquid crystal phases at the many-body level (see Fig. \ref{fig:liquidcrystal}). We identify three relevant time scales in this system: the frequency of the vibrational mode of the dimers $\nu$, the optomechanical damping rate $\gamma_{\rm opt}$ and the rate of thermalization $\gamma$. For limiting cases we present analytic results on the ordering of individual dimers and numerical results on the non-equilibrium many-body dynamics.

\textit{Model -} We consider a system of $N/2$ dimers consisting of $N$ nanospheres, trapped inside and optical cavity. The Hamiltonian is decomposed as $H=H_{\rm sys}+H_{\rm om}$ with
\begin{eqnarray}
\label{sysham}
H_{\rm sys}=\sum_{j=1}^{N}\left(\frac{\mathbf{p}_{j}^2}{2m}+V_{\rm trap}(\mathbf{x}_{j})\right)+\qquad\qquad\qquad\\
\sum_{j=1}^{N/2}\frac{m\nu^{2}}{2}(|\mathbf{x}_{2j-1}-\mathbf{x}_{2j}|-x_{0})^2+\Gamma_{0}\sum_{i\neq j}V_{\rm int}(|\mathbf{x}_{i}-\mathbf{x}_{j}|)\nonumber\;,
\end{eqnarray}
the system Hamiltonian. Here, $m$ is the mass of the nanospheres, $\mathbf{x}_{j}$ and $\mathbf{p}_{j}$ are the position and momentum of particle $j$, respectively, $V_{\rm{trap}}$ is the trapping potential, $\nu$ the frequency of the vibrational mode of the dimers, $x_{0}$ is the equilibrium separation and $V_{\rm int}$ is the direct dipolar pair interaction, which can be tuned by the parameter $\Gamma_{0}$ \cite{MARET}. In a frame, rotating with the laser drive, the optomechanical Hamiltonian is given by
\begin{equation}
\label{optham}
H_{\rm om}=-\Delta(\mathbf{x}_{1}\ ...\ \mathbf{x}_{j})|a|^2+\Omega(a+a^{\ast}) + V_{\rm c}\;,
\end{equation}
where $a(t)$ [$\sqrt{Js}$] is a normal variable, which describes the dynamics of the optical mode, $\Omega$ [$\sqrt{J/s}$] characterizes the drive strength, $V_{\rm c}$ is a second compensating potential, and $\Delta=\omega_{d}-\omega(\mathbf{x}_{1}\ ...\ \mathbf{x}_{j})$ is the detuning of the laser drive from the optical resonance. The interaction derives from the electric polarizability $\alpha_{\rm p}$ of the nanospheres \cite{CHANGZOLLER,CIRAC}, which, for sub-wavelength particles, gives rise to a position-depended cavity resonance frequency $\omega(\mathbf{x}_{1}\ ...\ \mathbf{x}_{j})=\omega_0-(gC/2)\sum_{j}|\mathbf{F}(\mathbf{x}_{j})|^2$, where $\omega_0$ is the bare cavity frequency. $\mathbf{F}(\mathbf{x})$ is the normalized mode function of the cavity mode, $C$ is the mode volume and $g=\alpha_{p} \omega_0/C$ is the optomechanical coupling strength, which is proportional to the volume of the particle. We assume that the cavity mode can be approximated by a standing wave $\mathbf{F}(\mathbf{x})\simeq \boldsymbol{\epsilon}\sqrt{2/C}\sin(kx)$, where $k$ is the wave number and $\boldsymbol{\epsilon}$ is the polarization in the $yz$ plane, so that $\Delta(\mathbf{x}_{1}\ ...\ \mathbf{x}_{j})=\Delta(x_{1}\ ...\ x_{j})$. 

The full dynamics of the optical part, including decay from the cavity, is described by the equations of motion for the optical amplitude $\dot{a}=(i\Delta(x_{1}\ ...\ x_{j})-\kappa)a+i\Omega$, where $\kappa$ is the field decay rate. The drive can be eliminated by the transformation $a\rightarrow\alpha +a'$ with $\alpha=\Omega/(\bar{\Delta}+i\kappa)$ the average amplitude and $\bar{\Delta}=(\omega_{d}-\omega)$ the detuning from the bare cavity resonance. The equation of motion for the transformed optical amplitude is $\dot{a}'=(i\bar{\Delta}-\kappa)a'+i(\Delta(x_{1}\ ...\ x_{j})-\bar{\Delta})(\alpha+a')$. Substitution of $a\rightarrow\alpha +a'$ in the first term in Eq. (\ref{optham}) gives rise to four terms. One of them corresponds to an effective optical potential along the cavity axis $V(x_{1}\ ...\ x_{j})=-|\alpha|^2\Delta(x_{1}\ ...\ x_{j})$, which can be compensated by $V_c$ of the second optical mode. For the choice of a mode separated by one free spectral range from $\omega$, the mode function in the focal range of the cavity is approximately given by $\mathbf{G}(x)\simeq\boldsymbol{\eta}\sqrt{2/C}\cos(kx)$, where $\boldsymbol{\eta}$ is the polarization. When $\boldsymbol{\epsilon}$ and $\boldsymbol{\eta}$ are orthogonal, the two modes do not interfere and the effective potentials add up, i.e. $V(x_{j})=-g(|\alpha_{\rm s}|^2\sin^{2}(kx_{j})+|\alpha_{\rm c}|^2\cos^{2}(kx_{j}))$. For $|\alpha_{\rm s}|=|\alpha_{\rm c}|$, the potential is independent of $x_{j}$ and the forces cancel. The second mode is driven on resonance and therefore does not lead to additional damping (or amplification).

We find the equations of motion of the nanosphere positions and momenta 
\begin{align}
\label{parteq}
\dot{\mathbf{x}}_{j}&=\mathbf{p}_{j}/m\nonumber\\
\dot{\mathbf{p}}_{j}&=-\frac{\partial H_{\rm sys}}{\partial\mathbf{x}_{j}}+\left(\alpha a^{\ast}+\alpha^{\ast}a+|a|^2\right)\frac{\partial\Delta(x_{1}\ ...\ x_{j})}{\partial x_{i}}\;.
\end{align}
The cavity decay rate $\kappa$ sets a finite time scale for the cavity to respond to changes in the particle positions so that the optomechanical feedback does not only depend on the particle positions, but also on the particle momenta along the cavity axis \cite{KIPPENBERG}. This gives rise to amplification  (for $\bar{\Delta}>0$) or damping (for $\bar{\Delta}<0$), respectively. Optomechanical cooling of nanospheres has been studied previously in the Lamb-Dicke regime \cite{CHANGZOLLER}, while here,  we focus on a different regime, similar to Ref. \cite{FREECOOLING}, in which the particles move almost freely and the relevant frequency stems from the modulation of the optical feedback force at $\omega_{\rm mod}=2kp/m$. The damping force has resonances for $\bar{\Delta}=\pm|\omega_{\rm mod}|$, while in the regime we are interested in, $|p| \ll m\bar{\Delta}/(2k)$,  the cooling rate is approximately constant
\begin{equation}
\gamma_{\textrm{opt}}\simeq\frac{2g^{2}|\alpha|^{2}k^{2}\bar{\Delta}\kappa}{m(\bar{\Delta}^2+\kappa^2)^2}\;.
\end{equation}
Finally, the equations of motion take the form of a modified Langevin equations with
\begin{align}
\label{langevin}
m\ddot{\mathbf{x}}=&-\frac{\partial H_{\rm sys}}{\partial\mathbf{x}}-m(\gamma+\gamma_{\rm opt})\dot{\mathbf{x}}+\xi_{x}\nonumber\\
m\ddot{\mathbf{y}}=&-\frac{\partial H_{\rm sys}}{\partial\mathbf{y}}-m\gamma\dot{\mathbf{y}}+\xi_{y}\;.
\end{align}
Here, $\gamma$ is the rate of thermalization and $\xi_{x}$ and $\xi_{y}$ are mutually uncorrelated Langevin forces, as characterized by $\langle\xi_{x,y}(t)\xi_{x,y}(t')\rangle=k_{B}Tm\gamma\delta(t-t')$. 

\begin{figure}[htb]
\centerline{\includegraphics[width=8.5cm]{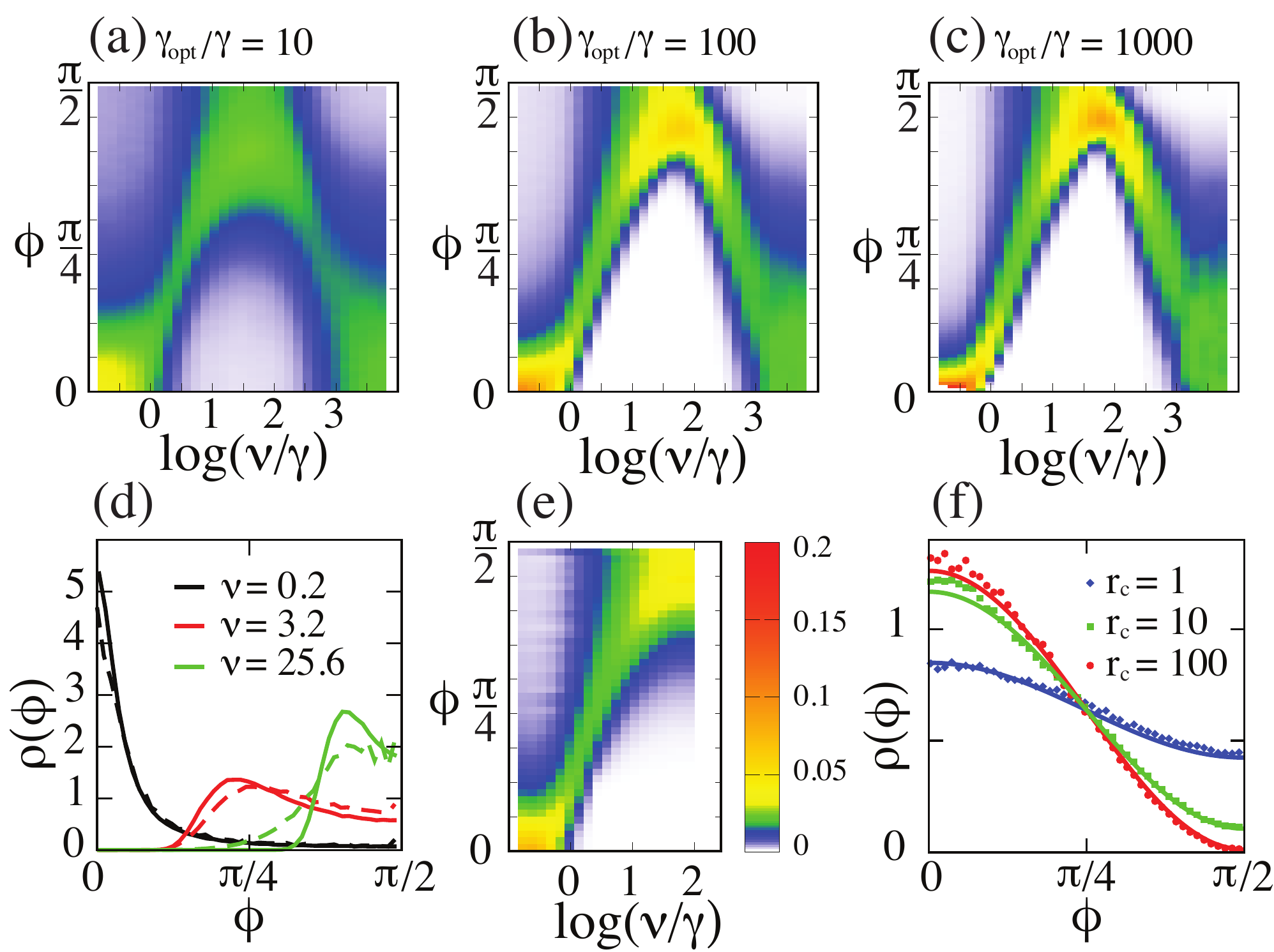}}
\caption{Steady-state probabilities of the orientation $\phi$ as a function of the dimer frequency from numerical integration of the second order Langevin equation Eqs. (\ref{langevin}) for cooling rate ratios $\gamma_{\rm opt}/\gamma = 10$ (a), $\gamma_{\rm opt}/\gamma = 100$ (b), and $\gamma_{\rm opt}/\gamma = 1000$ (c). The comparison of the analytic expressions Eq. (\ref{Pxy}) for small frequencies  (d dashed lines) and numerical results (d full lines) show good agreement over a wide range of frequencies. (e) Analytic results from Eq. (\ref{Pxy}) for a range of frequencies with $\gamma_{\rm opt}/\gamma = 100$ predicts the transition from $\phi=0$ orientation to $\phi=\pi/2$. For larger frequencies the model breaks down as it predicts $\phi=\pi/2$ for large $\nu$. (e). In the rigid-rotor regime, the analytic approximation Eq. (\ref{rigrot}) is in agreement with the numerical simulation as shown for various cooling rates $r_c = \gamma_{\rm opt}/\gamma$ (f).}
\label{fig:comparison}
\end{figure}
\textit{Results -} We first focus on the dynamics of a single dimer described by Eq. (\ref{langevin}). Separating the dynamics into the trivial center-of-mass motion and the relative coordinates $(\mathbf{x}_{1}-\mathbf{x}_{2})/\sqrt{2}\equiv(x,y)$, and to the extent that optomechanical coupling between the particles can be neglected, the non-linear force is $\frac{\partial H_{\rm sys}}{\partial \mathbf{r}} = - \mathbf{r} m\nu^2 \left(1-({2 x_{0}/(x^2+y^2)})^{1/2}\right)$, where $\mathbf{r} = (x,y)$ and $x_0$ is the dimer separation. Fig. \ref{fig:comparison} (a-c) show the steady state solutions of the orientation of single dimers as a function of the dimer frequency for various cooling rates from numerical integration of Eq. (\ref{langevin}). Remarkably, when exposed to uni-directional friction, a loosely connected dimer as well as a rigid rotor tends to align orthogonal to the direction of friction, whereas a dimer of moderate stiffness aligns parallel to it. This can be understood from three competing effects which derive from the order of the relevant time scales in the system (see Fig. \ref{fig:illustration}c): $\gamma^{-1}$, $\gamma_{\rm opt}^{-1}$ and $\nu^{-1}$. While $\gamma$ and $\gamma_{\rm opt}$ set the scales of thermalization and non-uniform friction, $\nu$ sets the time scale at which the degrees of freedom mix due to the non-linear nature of the force term. Assuming that $\gamma_{\rm opt}>\gamma$, there are three limiting parameter regimes: (i) $\gamma_{\rm opt}\gg\gamma\gg\nu$, (ii) $\gamma_{\rm opt}\gg\nu\gg\gamma$ and (iii) $\nu\gg\gamma\gg\gamma_{\rm opt}$. In the following we analytically study these limiting cases and give an intuitive explanation of this remarkable non-equilibrium ordering phenomenon. 

In (i) and (ii), $\gamma_{\rm opt}$ is the largest scale. Due to the resulting separation of time scales of the motion in the $x$ and $y$ directions, the steady-state distribution is of the form $P(x,y)=P(y|x)P(x)$. Here, $P(y|x)$ is the normalized steady-state distribution of the fast $y$ direction given a fixed $x$. Consequently, the average energy $\langle V\rangle_{x}= \int_{-\infty}^{\infty}V(x,y)P(y|x)$ determines the distribution $P(x)$ for the $x$ direction via a Markov process with the scaled temperature $T_{x}=T \gamma/(\gamma+\gamma_{\rm opt})$, thus 
\begin{equation}
\label{Pxy}
P(x,y)=\frac{e^{-V(x,y)/(k_{B}T)}}{\int_{-\infty}^{\infty} dy\; e^{- V(x,y)/(k_{B}T)}}\frac{e^{-\langle V\rangle_{x}/(k_{B}T_{x})}}{\int_{-\infty}^{\infty}dx\; e^{-\langle V\rangle_{x}/(k_{B}T_{x})}}\;.
\end{equation}
In regime (i), the spatial fluctuations are much larger than $x_{0}$, so that $V(x,y)\simeq m \nu^2(x^2+y^2)/2$ and the integrals can be evaluated analytically. The distribution $P(\phi)=\int_{0}^{\infty}rdr P(r\cos\phi,r\sin\phi)$ for the orientation of the dimer $\phi$ is found as
\begin{equation}
\label{geom}
P(\phi)=\frac{\sqrt{\gamma(\gamma+\gamma_{\rm opt})}}{2\pi((\gamma+\gamma_{\rm opt})\cos^{2}\phi+\gamma\sin^{2}\phi)}
\end{equation}
with the maximum for $\phi=0$. In case (ii), the harmonic approximation of $V(x,y)$ breaks down, and Eq. (\ref{Pxy}) is evaluated numerically, with the results shown in Fig.\ref{fig:comparison}(d and e) which are in agreement with numerical integration of Eq. (\ref{langevin}) shown in Fig.\ref{fig:comparison}(b).

In the rigid-rotor regime (iii), motion in the radial direction is suppressed, so that $\sqrt{x^{2}(t)+y^{2}(t)}=x_{0}$ and we can derive a Langevin equation for $\phi$ alone: $mx_{0}\ddot{\phi}=-m(\gamma+\gamma_{\rm opt}\cos^{2}\phi)\dot{\phi}+\xi_{\phi}$ where $\xi_{\phi}=\xi_{x}\sin\phi+\xi_{y}\cos\phi$ so that $\langle\xi_{\phi}(t)\xi_{\phi}(t')\rangle=k_{B}Tm\gamma\delta(t-t')$ is independent of $\phi$. This equation describes thermal motion of the orientation of the dimer with angle-dependent damping. Since there is no conservative force, the motion is overdamped, so that $\ddot{\phi}\simeq\gamma(\phi)\dot{\phi}$, where $\gamma(\phi)=\gamma+\gamma_{\rm opt}\cos^{2}\phi$, and the Fokker-Planck equation reduces to $\partial P/\partial t=(\gamma+\gamma_{\rm opt}\cos^{2}\phi)^{-1} \partial P/\partial \phi$.
The stationary solution is
\begin{equation}
\label{rigrot}
P(\phi)=\frac{\gamma+\gamma_{\rm opt}\cos^{2}\phi}{2\pi\gamma+\pi\gamma_{\rm opt}}
\end{equation}
and has a maximum at $\phi=0$.

This motivates the following intuitive picture: The dominant mechanism in (i) is purely geometrical. Here, the distribution $P(x,y)$ is a Gaussian that is squeezed in the direction of cooling $x$ and, therefore, the most likely orientation is $\phi=0$. In the rigid-rotor case (iii), in which no orientation-dependent energies are involved, the anisotropy of the steady-state orientation is due to a purely dynamical effect. In this case, the dimer is dynamically attracted to orientations for which the fluctuations in the angular direction are suppressed. The opposite orientation is reached in the intermediate regime (ii), when a third, purely energetical effect, is dominant. The motion is mostly confined to configurations of constant $x$, for which the potential energy changes from a double harmonic well at $x=0$ to a single anharmonic well ($V\propto x^{4}$) at $x=\pm x_{0}$. With the average kinetic energy $\langle T\rangle$ constant and using the virial theorem $\langle V \rangle = 2/n \langle T \rangle$, with $n$ the power of the external potential, we find that the probability $\exp[-\beta_x \langle V \rangle]$ is largest for $\phi=\pi/2$.

Let us consider the experimental feasibility in a configuration as shown in Fig.~\ref{fig:illustration}. The confinement of the dumbbells to the xy plane can be provided by an external standing wave optical trap crossing the Fabry-Perot cavity in $z$-direction. The three relevant time scales can be controlled over a large range of parameters: Dimer frequencies up to $\nu\simeq 2\pi\times 1\ \textrm{kHz}$ can be reached e.g. with spring constant $k \simeq 0.2\ \textrm{pN}/\mu m$ of DNA and silica nanospheres with a radius of $r = 50\ \textrm{nm}$ (mass $m=1.2\times10^{-18}\ {\rm kg}$). We note that spring constants smaller by orders of magnitude are possible above the persistence length of $50nm$ \cite{BUSTAMANTE2}. The optical damping is provided by a cavity with length $L\simeq 10^{-2}\ {\rm m}$ and mode waist of $w\simeq 10^{-4}\ {\rm m}$. We find the optomechanical coupling $g\simeq 2\pi\times 10^4\ {\rm Hz}$ via the mode volume $C=(\pi/4)Lw^{2}$. We further assume a cavity finesse of $\mathcal{F}\simeq 10^{5}$, so that $\kappa\simeq 2 \times 10^{5}\ {\rm Hz}$ at a wavelength of $\lambda=1064\ {\rm nm}$. When we further choose $\bar{\Delta}\gtrsim 5\kappa$, a power for the cooling laser of $P_{\rm drive}=3 \times 10^{-5} \ {\rm W}$ results in a cooling rate of $\gamma_{\rm opt}\lesssim 2\pi\times 4\ {\rm Hz}$. For the thermal environment we assume room temperature $T=293\ {\rm K}$ and $\gamma=0.05\ {\rm Hz}$, which corresponds for the chosen nanospheres to an air pressure of approximately $10^{-5} \ {\rm mbar}$. Therefore, rates of up to $\gamma_{\rm opt}/\gamma \approx 100$ are possible. Note that even higher rates have recently been achieved experimentally by optical feedback cooling \cite{FEEDBACK}.

\textit{Many-body phases -} Liquid crystal phases of dimers have been previously studied in equilibrium and non-equilibrium \cite{SOFTMATTERBOOK}. Non-isotropic friction and the resulting ordering may offer novel tools to guide the self-assembly towards preferred structures and to study novel nematic phases. We numerically study the system described by the Eq. (\ref{sysham}) with the experimental parameters as given above. Fig. \ref{fig:liquidcrystal}(a) depicts an ensemble of dimers without optical friction. For this choice of parameters, the system is in the liquid phase and the dimer orientations are distributed uniformly. Additional non-isotropic cooling with $\gamma_{\rm opt}/\gamma = 100$ induces a non-equilibrium transition to a phase characterized by the single- and many-dimer order parameters respectively shown in Fig. \ref{fig:liquidcrystal} (b) and (c). The single dimers are still aligned. In addition, the interplay between orientation and many-body dynamics leads to a remarkable phase with liquid order in the $y$-direction and solid order in the $x$-direction. With the frequency $\nu' = 500 \nu$ and keeping all other parameters fixed, a different pattern with all individual dimers reorientated orthogonal to the direction of cooling (Fig. \ref{fig:liquidcrystal}(c)) is found. Again the many-body dynamics leads to ordering along the $x$-direction, as measured by the directional pair-correlation function $g_{x,y}(x) = \langle \delta((x_i - x_j) - x) \rangle$ depicted in Fig. \ref{fig:liquidcrystal}(c). The patterns in Fig. \ref{fig:liquidcrystal} (b) and (c) are non-equilibrium analogs of liquid crystal phases in equilibrium. 
For large $\Gamma_0$ above the melting temperature \cite{MARET}, two-dimensional dipolar particles self-assemble into a triangular lattice. The guided orientation of individual dimers induced by non-isotropic friction is competing with the orientation in the triangular lattice in which the dimer orientations are random integer multiples of $\pi/3$. This frequency-dependent ordering can be used to create patterns of dimers with different orientations. Fig. \ref{fig:liquidcrystal}(d) depicts a mixture of dimers with $\nu_1=500 \nu$ and with $\nu_2=\nu$. In the absence of direct interactions these species order with almost orthogonal relative alignment. Increasing the interactions $\Gamma_0 > \Gamma_{\rm melt}$ above the critical melting temperature, the system forms a crystal with triangular ordering and separate orientational order of the two individual dimer species. 

In summary, we have presented a realistic optomechanical setup that allows for novel dissipative control over the orientation of dimers composed of nanospheres. We have shown that this approach can be used to prepare non-equilibrium analogs of liquid crystals and to study transitions in mixtures of multiple species of dimers. The only relevant parameters that determine the non-equilibrium ordering are the time scales of the vibrational mode of the dimers and the rates of thermalization and non-isotropic friction. Complex structures of DNA-connected nanospheres are of growing interest, and we hope that the approach discussed here will provide useful means to optical control of such systems, complementary to direct optical manipulation via optical tweezers. The presented mechanism does not rely on any specific properties of nanospheres and applies, at least in principle, in general to complex structures of dielectric objects, such as viruses and bacteria \cite{VIRUS}. We speculate that, in the longer run, it may also be applied to complex molecules, and may even prove fruitful as a microseeding technique for the nucleation of complex molecules. 

\textit{Acknowledgments.} We thank P.~Rabl, C.W.~Gardiner, M.~Gr\"unwald, and C.~Dellago for fruitful discussions. Work at Innsbruck is supported by the integrated project AQUTE, the Austrian Science Fund through SFB F40 FOQUS, and by the Institut f\"ur Quanteninformation.

\bibliographystyle{prsty}

\end{document}